\documentclass[10pt,twocolumn]{article}
\usepackage{graphicx} 
\graphicspath{{figures}} 
\usepackage[font=small,labelfont=bf]{caption} 

\usepackage{xcolor}
\definecolor{black}{HTML}{212427}
\definecolor{blue}{HTML}{0563C1}
\definecolor{red}{HTML}{B51700}

\usepackage{hyperref,nameref}
\hypersetup{colorlinks=true,allcolors=blue}
\newcommand{\rref}[2]{\hyperref[#1]{\ref{#1}#2}} 

\color{black}
\usepackage{setspace} 
\usepackage[margin=0.67in]{geometry} 
\usepackage[switch]{lineno} 
\usepackage{anyfontsize}

\usepackage{fancyhdr,lastpage}
\usepackage{titlesec}
\titleformat{\section}{\Large\bfseries}{}{0mm}{}
\titleformat{\subsection}{\bfseries}{}{0mm}{}
\titlespacing{\section}{0pt}{\baselineskip}{0pt}
\titlespacing*{\section}{0pt}{\baselineskip}{0pt}
\titlespacing*{\subsection}{0pt}{\baselineskip}{0pt}

\usepackage{amsmath,amssymb}
\newcommand{\Ang}[0]{\mathring{\mathrm{A}}} 
\renewcommand{\t}[1]{\text{#1}} 

\usepackage{multirow} 

\usepackage{soul} 
\setstcolor{red}
\setul{}{0.7pt}

\usepackage[
  backend=biber,        
  autocite=superscript, 
  sorting=none,         
  style=numeric-comp,   
  maxnames=100,         
  minnames=100,         
  giveninits=false,     
  autopunct=false,      
  doi=true,             
  hyperref=true         
]{biblatex}
\DeclareFieldFormat{labelnumberwidth}{\;[#1]\!\!\!\!} 
\renewbibmacro{in:}{} 
\AtNextBibliography{\small} 
\AtEveryBibitem{\clearfield{number}} 
\AtEveryBibitem{\clearfield{month}} 

\begin{document}

\twocolumn[
  \begin{center}
    \large
     \textbf{Precipitation strengthening as a collective multi-dislocation phenomenon}
  \end{center}
  Mahmudul Islam$^{1,2}$, Jason Ogbebor$^1$, Nicolas Bertin$^2$, Sylvie Aubry$^2$, Vasily V. Bulatov$^2$, and
  Rodrigo Freitas$^1${\footnotemark[1]} \\
  $^1$\textit{\small Department of Materials Science and Engineering, Massachusetts Institute of Technology, Cambridge, MA, USA} \\
  $^2$\textit{\small Lawrence Livermore National Laboratory, Livermore, CA, USA} \\

  {\small Dated: \today}
  
  \vspace{-0.15cm}
  \begin{center}
    \textbf{Abstract}
  \end{center}
  \vspace{-0.35cm}
  
Precipitation strengthening is a cornerstone of physical metallurgy, delivering otherwise unattainable combinations of strength and ductility. The approach relies on nanoscale precipitates that impede the motion of dislocations, the primary carriers of metal plasticity. Historically, precipitation strengthening has been rationalized via two idealized, limiting mechanisms: dislocations either cut through or bow around precipitates. However, under many technologically relevant conditions, multiple dislocations can interact with a precipitate at the same time, potentially giving rise to distinct precipitate-dislocation dynamics that in situ experiments are yet to fully resolve. Here, using large-scale atomistic simulations ($\sim10^{8}$ atoms) that capture these dynamics, we demonstrate that in this regime, the cutting-versus-bowing dichotomy is incomplete: strengthening arises as an emergent phenomenon driven by concurrent multi-dislocation interactions. These interactions simultaneously induce dislocation accumulation at interfaces, storage within precipitates, and precipitate-mediated multiplication inside the matrix. Our findings establish a distinct mechanistic regime of precipitation strengthening beyond traditional models and provide a quantitative foundation for predicting its behavior.

  \vspace{0.4cm}
]
{
  \footnotetext[1]{Corresponding author (\texttt{rodrigof@mit.edu}).}
}


\section{Introduction}

\noindent Metals are strengthened by precipitating nanoscale secondary-phase particles into a supersaturated matrix, a process known as precipitation strengthening\autocite{jiang_ultrastrong_2017,sun_precipitation_2019,tang_precipitation_2019,jang_shear_2021}. Widely regarded as a central nanoscale design strategy in modern metallurgy, this approach predates nanoscience by many decades, with its empirical discovery\autocite{wilm1911physical} in 1906 preceding the experimental tools needed to directly visualize nanoscale precipitates. Precipitates strengthen metals by impeding the motion of dislocations, curvilinear defects that mediate plastic deformation. This basic principle suggests that precipitates that more effectively obstruct dislocation motion should produce greater strengthening. Yet a quantitative connection between atomic-scale dislocation-precipitate interactions and macroscopic strength remains unresolved\autocite{cui_multiscale_2024}.

Established theories of precipitation strengthening\autocite{kelly_strengthening_1971,reppich_new_1982,ardell_precipitation_1985,argon_strengthening_2007} rest on an idealized premise: an encountering dislocation either shears through a precipitate (cutting) or bypasses it (Orowan looping), depending on which path is energetically favored. This binary framework, illustrated in fig.~\rref{figure_1}{a}, has inspired numerous analytical models\autocite{cui_multiscale_2024,kelly_strengthening_1971,nembach_particle_1997}. While these formulations capture specific limiting regimes\autocite{orowan1948symposium,jiang_ultrastrong_2017,ardell_precipitation_1985,seidman_precipitation_2002,wang_shearing_2024,takahashi_computational_2008}, they often deviate substantially from empirical observations. As shown in fig.~\rref{figure_1}{b}, comparisons with experimental data often reveal pronounced quantitative discrepancies, particularly near the peak strength. To reconcile such discrepancies, phenomenological fitting coefficients are introduced\autocite{cui_multiscale_2024,ardell_precipitation_1985,seidman_precipitation_2002, krug_comparison_2014}, a remedy that suggests the idealized framework omits some of the underlying physics. Furthermore, while classical theories predict an abrupt transition from shearing to bypassing, experiments consistently reveal a broad, gradual crossover (fig.~\rref{figure_1}{b}). Although precipitate size polydispersity is often invoked to reconcile this inconsistency\autocite{seidman_precipitation_2002,fang_statistical_2019,glazer_effect_1988,van_dalen_microstructural_2011}, this explanation presumes that the underlying transition is itself sharp, an assumption that has not been tested. These discrepancies may become increasingly consequential in heavily worked or additively manufactured metals, where high dislocation densities are likely to introduce interactions beyond those represented by the idealized single-dislocation model in fig.~\rref{figure_1}{a}. Collectively, these discrepancies underscore a gap in our mechanistic understanding of precipitation strengthening --- a framework that, despite its limitations, continues to direct modern alloy design and processing strategies.

Here, we show that precipitation strengthening can arise from collective interactions between multiple dislocations and precipitates, a setting not captured by the single-dislocation model. Using ultra-large-scale molecular dynamics (MD) simulations ($\sim\!10^8$ atoms) that resolve these processes in atomistic detail, we directly observe how multiple dislocations engage with precipitates during plastic deformation. In this regime, strengthening response cannot be reduced to a binary choice between shearing and bypass. Instead, it arises from the coupled development of dislocation accumulation at precipitate interfaces, dislocation storage within precipitates, and precipitate-mediated dislocation multiplication in the surrounding matrix. These processes occur concurrently and together control the precipitate contribution to strength, providing a microscopic basis for multi-dislocation precipitation hardening that extends beyond the traditional shearing-versus-bypass picture.


\section{Results} 

\subsection{Multi-dislocation interaction with precipitates}
\label{section_1}

\begin{figure*}[!tb]
  \centering
\includegraphics[width=\textwidth]{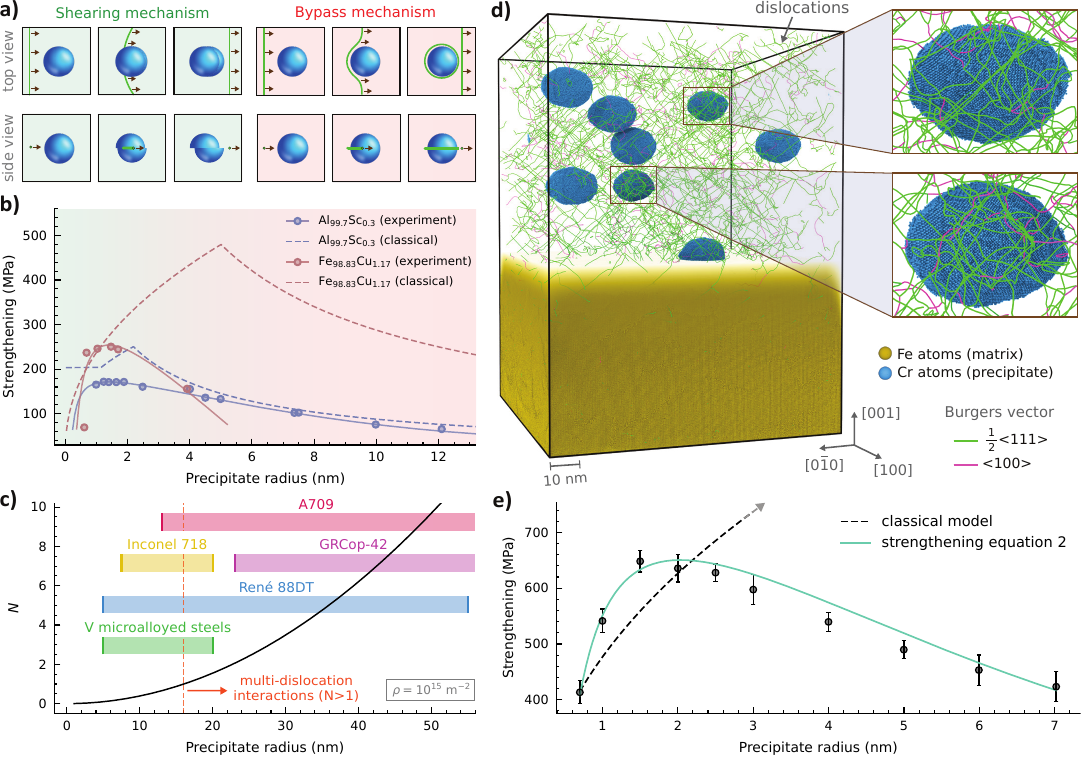}
  \caption{\label{figure_1}
\textbf{From the classical single-dislocation picture to multi-dislocation precipitate strengthening.}
\textbf{a)} Schematic of the classical single-dislocation picture: a dislocation either shears a precipitate or bypasses by bowing around it and leaving behind a closed loop.
\textbf{b)} Experimental strengthening data for Al-Sc\autocite{seidman_precipitation_2002} and Fe-Cu\autocite{osamura_precipitation_1994} compared with single-dislocation model predictions (described in Supplementary Information Note 1). The solid line is drawn as a guide to the eye. \textbf{c)} Number of dislocations $N$ within a precipitate interaction region as a function of precipitate radius ($R$) for a dislocation density of $10^{15}\,\t{m}^{-2}$. Over much of the precipitate size range relevant to engineering alloys\autocite{mohale_microstructural_2022,seltzman_precipitate_2023,tiley_coarsening_2009,baker_processes_2009,lee_evaluation_2023, lee_precipitate_2025}, $N > 1$, indicating that multi-dislocation interactions beyond the single-dislocation picture of \textbf{a)} are likely. \textbf{d)} Representative atomistic configuration of Fe containing spherical, monodisperse Cr precipitates ($R=7\,\t{nm}$, volume fraction $v_\t{f}=1\%$) during uniaxial compression at room temperature. 
\textbf{e)} Precipitation strengthening $\Delta \sigma$, as a function of $R$ at $v_\t{f} = 15\%$, where $\Delta \sigma$ is the increase in flow stress relative to the matrix reference. The dashed line shows the corresponding classical single-dislocation prediction; the grey arrow indicates that it continues to increase monotonically over the precipitate-size range shown. Solid line shows the prediction of partitioning-based strengthening relation (eq. \ref{eq:unified_strength}). Error bars denote
correlation-corrected standard errors of the mean steady-state flow stress.}
\end{figure*}

As summarized in Supplementary Information Note~1, classical models of precipitation strengthening are formulated in terms of isolated interactions between single dislocations and precipitates. A simple geometric 
estimate of the number of dislocations ($N$) within a precipitate's interaction region (see Methods section ``\nameref{method:count}'') shows that this regime is not representative of many commercial alloys. When $N > 1$, individual precipitates are likely to be encountered by multiple dislocations simultaneously rather than one dislocation at a time. Figure~\rref{figure_1}{c} illustrates this condition for a dislocation density of $10^{15}\,\t{m}^{-2}$, which can be reached during cold working\autocite{williamson_iii_1956}. This multi-dislocation regime spans much of the range of precipitate sizes and volume fractions used in practical alloy design and yet it falls outside the single-dislocation picture of fig.~\rref{figure_1}{a}. This disconnect motivates our explicit examination of multi-dislocation interactions with precipitates.

To directly probe this multi-dislocation regime, we carried out large-scale MD simulations of body-centered cubic (BCC) Fe single crystals containing coherent Cr precipitates, compressed along the [001] direction. In the series of simulations reported here, both the precipitate radius ($R$) and volume fraction ($v_\t{f}$) were varied; however, all precipitates within a given sample had the same radius, yielding a monodisperse microstructure. Elemental BCC Cr is elastically stiffer than the BCC Fe matrix, and because Cr atoms are only slightly larger than Fe atoms, this combination provides a well-defined mechanical contrast for precipitation strengthening while retaining precipitate coherency. All samples were deformed at room temperature over a range of strain rates. Figure~\rref{figure_1}{d} shows a representative snapshot of these samples.

As deformation proceeds, dislocations multiply and organize into a dense network that spans the microstructure\autocite{zepeda-ruiz_atomistic_2021,islam_dislocation-mediated_2025, islam_nonequilibrium_2025} (see Supplementary Information Movie~1 for the full temporal evolution of dislocation networks). In the representative configuration shown in fig.~\rref{figure_1}{d}, multiple dislocations impinge simultaneously on the same precipitate, forming transient multi-dislocation interaction zones (see Supplementary Information Note~2 for additional representative configurations). These configurations, which correspond to $N>1$ in fig.~\rref{figure_1}{c}, demonstrate that multi-dislocation interactions characterize the deformation under these conditions, rather than isolated single-dislocation events (see also Supplementary Information Note 13).

Across all samples, the stress-strain response is qualitatively similar: an initial linear elastic regime, followed by yielding as dislocations begin to move and multiply, and eventual convergence to steady-state flow at constant stress. Previously reported for pure metals\autocite{zepeda-ruiz_atomistic_2021} and solid solutions\autocite{islam_nonequilibrium_2025,islam_dislocation-mediated_2025}, this steady-flow regime coincides with saturation of the dislocation density, at which point the dislocation network attains a statistically stationary configuration. Here, we further observe that dislocation-precipitate interactions also become statistically stationary. Representative examples of stress-strain response and evolution of dislocation density are provided in Supplementary Information Notes~3-4.

The strengthening contribution of the precipitates is reflected in the flow stress. We express this contribution as the precipitation strengthening $\Delta \sigma$, defined as the increase in flow stress relative to that of the matrix under the same deformation conditions. Figure~\rref{figure_1}{e} shows $\Delta \sigma$ as a function of precipitate radius $R$ at $v_\t{f} = 15\%$. Consistent with the experimental trends in fig.~\rref{figure_1}{b}, $\Delta \sigma$ varies non-monotonically with $R$: it increases with $R$, reaches a maximum at $R \approx 2\,\t{nm}$, and then decreases for larger precipitates. In contrast, the classical models (dashed line in fig.~\rref{figure_1}{e}) predict strengthening to increase monotonically up to the cutting-to-bypass transition at $R\approx11.8\,\t{nm}$. Notably, the crossover from increasing to decreasing $\Delta \sigma$ is gradual rather than abrupt, despite the monodisperse precipitate microstructure. This behavior contrasts with the sharp shearing-to-bypass transition expected from classical models (fig.~\rref{figure_1}{b}) which is often attributed to precipitate-size polydispersity\autocite{seidman_precipitation_2002, fang_statistical_2019, glazer_effect_1988,van_dalen_microstructural_2011}. Instead, the smooth crossover emerges intrinsically from multi-dislocation interactions with precipitates. The same behavior is observed across all volume fractions investigated (see Supplementary Information Figure~4e).

\subsection{Dislocation partitioning by precipitates}
\label{section_2}

\begin{figure*}[!tb]
\centering
\includegraphics[width=\textwidth]{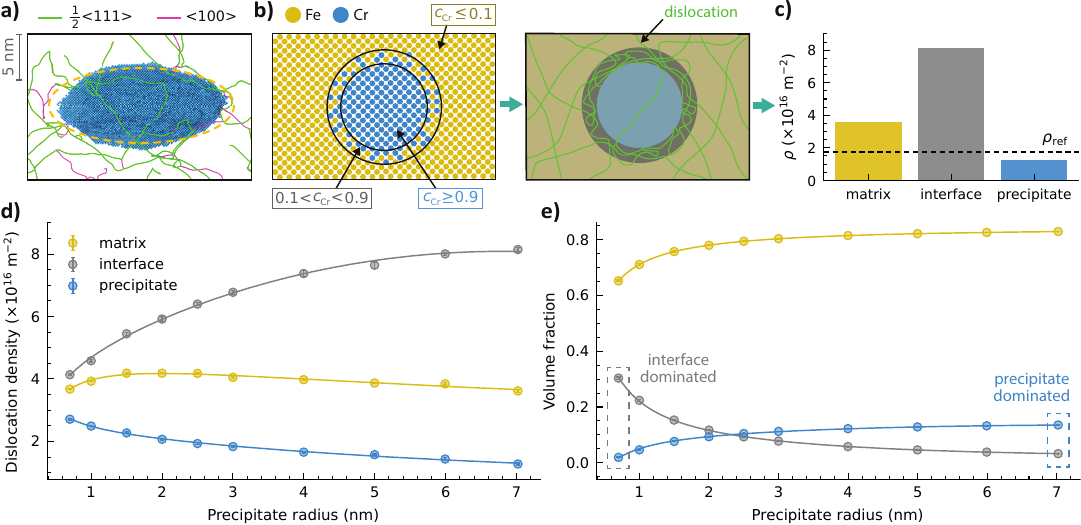}
\caption{\label{figure_2}
\textbf{Dislocation partitioning between matrix, precipitate, and interface.}
\textbf{a)} A typical precipitate after uniaxial compression to $\varepsilon=0.6$, illustrating precipitate shape change under loading. The dashed outline shows expected precipitate shape if it were to deform affinely with the matrix.
\textbf{b)} Composition-based voxel segmentation used to define matrix ($c_{\t{Cr}} \leq 0.1$), interface ($0.1 < c_{\t{Cr}} < 0.9$), and the precipitate ($c_{\t{Cr}} \geq 0.9$).
\textbf{c)} Dislocation density $\rho$ in the matrix, the precipitates, and their interfaces for $R = 7\,\t{nm}$ and $v_\t{f} = 15\%$. $\rho_\t{ref}$ is the dislocation density in pure Fe deformed under the same condition.
\textbf{d)} Dislocation densities in the matrix, at the interface, and within the precipitate as functions of precipitate radius $R$ (for $v_\t{f} = 15\%$).
\textbf{e)} Volume fractions of the matrix, interface, and within the precipitate as functions of $R$ (for $v_\t{f} = 15\%$). Error bars (smaller than the marker size and therefore not visible) indicate correlation-corrected standard errors of the mean of steady-state dislocation densities and microstructure fractions.}
\end{figure*}

Building on the observation that deformation involves concurrent interactions between multiple dislocations and a single precipitate, we now examine how dislocations partition between the matrix, the precipitates, and their interfaces. Figure~\rref{figure_2}{a} presents a magnified view of a precipitate of initial radius $R = 7\,\t{nm}$ after plastic straining of the sample to $\varepsilon = 0.6$. During deformation, the precipitate does not retain its initial spherical shape; instead, it evolves into an oblate ellipsoid with its minor axis aligned with the loading direction ([001]). This shape change is entirely due to dislocation shearing, as the precipitate remains coherent throughout deformation, with no evidence of phase transformation or deformation twinning at any stage (see Supplementary Information Note~5). The axial strain of the precipitate along the loading direction in fig.~\rref{figure_2}{a} is $\bar{\varepsilon}_\t{prec} = 0.5$ (see Supplementary Information Note~6), which is notably lower than the applied strain of $\varepsilon = 0.6$. This strain mismatch indicates limited dislocation transmission across the precipitate-matrix interface: although dislocations intermittently penetrate and shear the precipitate, a substantial fraction of them are arrested at the interface, while others are redirected through the surrounding matrix due to the elastic and lattice mismatch between Fe and Cr\autocite{ardell_precipitation_1985}.

To quantify how the competition between interfacial transmission and arrest redistributes dislocations, we resolve the dislocation density separately in the matrix, precipitates, and their interfaces (fig.~\rref{figure_2}{b}; see Methods section ``\nameref{method:dxa}'' for details). Figure~\rref{figure_2}{c} shows that the interfaces carry the highest dislocation density, indicating dislocation accumulation near the precipitates\autocite{zhou_designing_2026}. By contrast, the precipitates exhibit a substantially lower dislocation density than the matrix, consistent with incomplete dislocation transmission into the precipitate and limited storage within it. More importantly, the dislocation density in the matrix itself exceeds that of a precipitate-free matrix deformed under exactly the same conditions, indicating precipitate-mediated excess dislocation in the matrix\autocite{peng_nanoscale_2020, PENG2023103710}. Multi-dislocation interactions at the precipitates therefore give rise to three concurrent processes: interfacial dislocation accumulation, dislocation storage within precipitates, and precipitate-assisted dislocation multiplication into the surrounding matrix. All three take place at every precipitate and together constitute the microscopic basis of precipitation strengthening under these conditions.

Figure~\rref{figure_2}{d} shows how the matrix dislocation density $\rho_{\t{mat}}$ varies with precipitate size $R$. Across the full range of radii, $\rho_{\t{mat}}$ remains $\sim115\%$ higher than that of the precipitate-free matrix ($\sim 1.75 \times 10^{16}$ $\t{m}^{-2}$). This substantially elevated matrix density arises because precipitates impose local constraints on dislocation motion (observed in Supplementary Information Movie~1), promoting line bowing\autocite{ardell_precipitation_1985}, line-length increase\autocite{nembach_particle_1997}, junction formation\autocite{bulatov_network_2025}, and the emission of additional dislocation segments into the surrounding matrix\autocite{peng_nanoscale_2020, PENG2023103710}. At a fixed precipitate volume fraction, $\rho_{\t{mat}}$ varies non-monotonically with precipitate size: it increases with $R$, reaches a maximum, and then decreases for larger precipitates. This trend reflects a balance between individual obstacle strength and dislocation-precipitate interaction frequency. Large precipitates are stronger obstacles, but at fixed $v_\t{f} = 15\%$ they are fewer in number and more widely spaced than small precipitates, reducing the frequency of encounters with moving dislocations. Small precipitates, by contrast, are more numerous but are more readily sheared, making them weaker obstacles. The maximum in $\rho_{\t{mat}}$ therefore occurs at $R\approx2\,\t{nm}$ when precipitates are large enough to obstruct dislocation motion but still numerous enough to interact frequently with the evolving dislocation network.

Superimposed on this elevated matrix-density background, both $\rho_{\t{int}}$ and $\rho_{\t{prec}}$ vary monotonically with precipitate size. For small precipitates, interfacial accumulation is limited ($\rho_\t{int} \approx 1.15\rho_\t{mat}$) while transmission into the precipitate is substantial ($\rho_\t{prec} \approx 0.75\rho_\t{mat}$). As $R$ increases, $\rho_\t{int}$ rises while $\rho_\t{prec}$ falls. Both trends have the same physical origin: larger precipitates offer greater intrinsic resistance to dislocation slip, causing more dislocations to be arrested at the interface and fewer transmit into the precipitate. Those arrested at the interface build up a back-stress that, together with the precipitate-matrix stiffness contrast, resists further dislocations approaching. Note, however, that transmission never vanishes. Even for the largest precipitates, $\rho_\t{prec} \approx 0.35\rho_\t{mat}$, at odds with complete obstacle bypass, e.g. Orowan looping (fig.~\rref{figure_1}{a}). Dislocation transmission and interfacial accumulation therefore remain concurrent across the full range of precipitate sizes and volume fractions studied here, rather than separating into distinct cutting and bypass regimes\autocite{chatterjee_discrete_2021, stewart_predicting_2024} (see Supplementary Information Figure~9).

Precipitate size controls not only how dislocations partition between the matrix, interface, and precipitate, but also the volume fractions occupied by these three regions. Figure~\rref{figure_2}{e} shows the volume fractions of the matrix ($f_{\t{mat}}$), interface ($f_{\t{int}}$), and precipitates ($f_{\t{prec}}$) as functions of $R$. For small precipitates, the interfacial region occupies a disproportionately large fraction of the microstructure. This arises partly from the geometric increase in the surface-to-volume ratio with decreasing $R$, and partly from shear-induced interface broadening and associated diffusion\autocite{kovarik_microtwinning_2009}, which affects smaller and more shearable precipitates more strongly (see Supplementary Information Figures~10e and 10f). As a result, $f_{\t{int}}$ can become comparable to, or even exceed, $f_{\t{prec}}$. As $R$ increases, however, the interfacial region occupies a progressively smaller fraction of the microstructure, becoming nearly negligible at the largest radii (see Supplementary Information Figure~10 for other volume fractions).

Taken together, these observations clarify how precipitate size controls multi-dislocation strengthening response (fig.~\rref{figure_1}{e}). At small $R$, precipitates offer little resistance to gliding dislocations, which readily transmit across the interfaces to shear the precipitates. Accordingly, $\rho_\t{int}$ barely exceeds $\rho_\t{mat}$. At large $R$, $\rho_\t{int}$ rises well above $\rho_\t{mat}$ and exerts a substantial back-stress, but strengthening is again limited because the interfacial regions occupy only a small fraction of the material volume. In the intermediate range of precipitate sizes, the precipitates become more resistant while their interfaces still occupy appreciable fractions of the microstructure, thus allowing precipitate shear resistance and interfacial back-stress to contribute concurrently. This same intermediate range also maximizes precipitate-mediated dislocation multiplication in the matrix ($\rho_\t{mat}$ in fig.~\rref{figure_2}{d}), which adds a third concurrent contribution to strengthening. The combined effect of these contributions therefore places the strengthening maximum at $R \approx 2\,\t{nm}$, which is well below the classical model prediction (fig.~\rref{figure_1}{e}). And because the dislocation partitioning and interior volume fractions all evolve continuously with $R$ (figs.~\rref{figure_2}{d} and \rref{figure_2}{e}), this maximum is smooth, with no need to invoke precipitate-size polydispersity.

\subsection{A partitioning-based strengthening relation}
\label{section_3}

\begin{figure}[!tb]
  \centering
\includegraphics[width=0.5\textwidth]{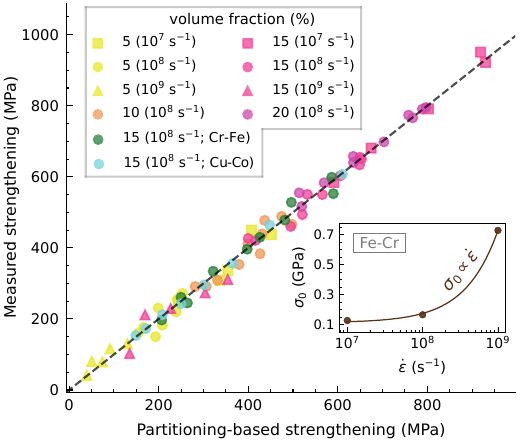}
  \caption{\label{figure_3}
  \textbf{A partitioning-based strengthening relation.} Strengthening from eq.~\ref{eq:unified_strength} compared with measurements across different $R$, $v_{\t{f}}$, and strain rates. Unless otherwise noted, the data correspond to the Fe-Cr system. There are 78 data points, each of which represents a sample with $\sim 10^8$ atoms. The dashed line denotes perfect agreement. The inset shows the fitted intrinsic friction stress $\sigma_0$ as a function of strain rate for the Fe-Cr system.}
\end{figure}

Having identified three distinct regions involved in precipitation strengthening, we now relate them quantitatively to the strengthening response. Consistent with the classical Taylor hardening equation\autocite{taylor_mechanism_1934}, which relates flow stress to the square root of dislocation density, Supplementary Information Figure~11a shows that precipitation strengthening closely co-varies with the total dislocation density. However, this Taylor-like relation becomes increasingly inaccurate as the precipitate volume fraction increases (Supplementary Information Figure~11b and 11c), indicating that total dislocation density alone does not fully capture this strengthening response. We therefore posit that, in multi-dislocation regime, plastic strength depends not only on how many dislocations are stored in the material, but also on where they are stored.

Recognizing that dislocations stored in the matrix, within precipitates, and at the interfaces obstruct dislocation motion to different degrees, we introduce a region-resolved Taylor-type relation for the stress contribution, in which $\sigma_A$ is the stress needed to move dislocations through region $A\in \{\t{mat},\t{int},\t{prec}\}$ against the dislocations already stored there:
\begin{equation} 
\sigma_{A} = M\alpha_{A}\mu_{A}b_{A}\sqrt{\rho_{A}}. \label{eq:region_taylor} 
\end{equation} 
Here, $\rho_A$ is the dislocation density measured in region $A$ (see Methods section ``\nameref{method:dxa}''), $\mu_A$ is the Voigt-averaged shear modulus, $b_A$ is the Burgers vector, $\alpha_A$ is the Taylor coefficient associated with that region\autocite{bulatov_network_2025}, and $M = 2.451$ is the standard Taylor factor converting stress resolved on the \{112\} glide planes to the applied flow stress for uniaxial deformation along the [001].

In a heterogeneous microstructure containing a matrix, interfaces, and precipitate interiors, the overall resistance to dislocation motion is limited by the least resistive deformation region. This is consistent with the observation in fig.~\rref{figure_2}{a} that these regions do not satisfy an iso-strain condition: the deformation accommodated by the matrix ($\varepsilon = 0.6$) and that by precipitates ($\bar{\varepsilon}_{p} = 0.5$) are measurably different. The three regions can therefore be reasonably assumed to serve as parallel deformation channels: if one region offers lower resistance and more readily accommodates plastic flow, dislocations will preferentially propagate through it, thereby limiting the net strengthening\autocite{zhu_heterostructured_2023}. This behavior can be expressed by the harmonic combination of the three resistances\autocite{picu_superposition_2010, LAGERPUSCH20003647, krug_comparison_2014}: 
\begin{equation}
  \sigma = \sigma_\t{0} + \left( \frac{f_{\t{mat}}}{\sigma_{\t{mat}}}+ \frac{f_{\t{int}}}{\sigma_{\t{int}}}+ \frac{f_{\t{prec}}}{\sigma_{\t{prec}}}\right)^{-1}. \label{eq:unified_strength}
\end{equation}
Here, $f_{\t{mat}}$, $f_{\t{int}}$, and $f_{\t{prec}}$ are the volume fractions of the matrix, interface, and precipitate regions, respectively, as defined in fig.~\rref{figure_2}{e}, and $\sigma_{0}$ is the friction stress due to the lattice (Peierls) resistance to dislocation motion. The strengthening, $\Delta\sigma$, is then obtained by subtracting the flow stress of the precipitate-free reference material from the predicted $\sigma$ obtained from eq.~\ref{eq:unified_strength}.

With a single set of fitted Taylor coefficients $\alpha_A$, equation~\ref{eq:unified_strength} captures strengthening measured in all 60 Fe-Cr samples over the entire range of $R$, $v_{\t{f}}$, and strain rates (fig.~\ref{figure_3}). The same coefficients, with no readjustment, produce the solid curve in fig.~\rref{figure_1}{e}. The agreement of this region-resolved Taylor model with the measurements is considerably closer than that of the conventional Taylor model relating flow stress to the total dislocation density (see Supplementary Information Figures~11b and 11c). Note also that the fitted parameters take physically sensible values. All three Taylor coefficients, $\alpha_{\t{mat}} = 0.254$, $\alpha_{\t{int}} = 0.199$, and $\alpha_{\t{prec}} = 0.302$, fall within the range reported for BCC metals\autocite{bulatov_network_2025}, and the friction stress $\sigma_0$ exhibits an approximately linear dependence on the strain rate (inset of fig.~\ref{figure_3}), consistent with behavior previously observed at high strain rates~\autocite{fan_strain_2021}.

So far, eq.~\ref{eq:unified_strength} has been tested on Fe-Cr alone, where dislocations accumulate preferentially at the interfaces and in the matrix. A more demanding test is a system in which dislocations distribute differently among the three regions. One way to produce such a system is to invert the stiffness contrast, i.e., embedding softer Fe precipitates in a stiffer Cr matrix (see Supplementary Information Note~9). In these samples, referred to here as Cr-Fe systems, $\rho_{\t{prec}}$ remains greater than $\rho_{\t{mat}}$ over the entire range of precipitate radii. The soft precipitates pose a low barrier to entry, so dislocations transmit into them freely and accumulate there, and the stored density $\rho_\t{prec}$ raises $\sigma_{\t{prec}}$ even though the precipitates themselves are easily sheared. Strengthening in these alloys therefore comes from storage within the precipitates, and the response no longer exhibits the maximum observed in fig.~\rref{figure_1}{e} (see Supplementary Information Figure 12b). Equation \ref{eq:unified_strength} captures this qualitative difference dependence on $R$ (fig.~\ref{figure_3}). It does so as well for the Cu-Co alloy (see Supplementary Information Note~10), which differs from Fe-Cr in crystal structure, chemistry, and dislocation character. The partitioning depicted in fig.~\ref{figure_2} and the strengthening relation built on it therefore carry over to multi-dislocation precipitation strengthening across crystal structures, chemistries, and phase contrasts involving coherent precipitates.

\begin{figure}[!tb]
\centering
\includegraphics[width=0.5\textwidth]{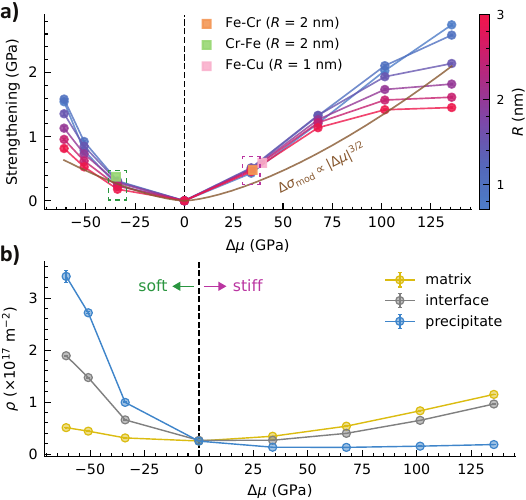}
\caption{\label{figure_4}
\textbf{Strengthening maps from alchemical modulus mismatch.}
\textbf{a)} Strengthening as a function of precipitate modulus mismatch $\Delta\mu =\mu_{\t{prec}}-\mu_{\t{mat}}$ and radius $R$ for alchemical precipitates at fixed $v_{\t{f}}=10\%$. The response is asymmetric about $\Delta\mu=0$, unlike the classical prediction, which scales as $|\Delta\mu|^{3/2}$. Soft and stiff precipitates with comparable $|\Delta\mu|$ therefore do not strengthen equivalently, as highlighted by the dashed boxes ($|\Delta \mu| \approx 34\,\t{GPa}$). Fe-Cr, Cr-Fe, and Fe-Cu samples ($v_{\t{f}}=10\%$) at the same conditions lie on the alchemical strengthening map at comparable modulus mismatch.
\textbf{b)} Steady-state dislocation densities in the matrix, interface, and precipitate regions as functions of $\Delta\mu$, for representative alchemical systems with $R=2\,\t{nm}$.}
\end{figure}

\subsection{Strengthening maps from stiffness mismatch}

The comparison between Fe-Cr and Cr-Fe points to the stiffness contrast between precipitate and matrix as the property that sets the dislocation partitioning: reversing its sign moved dislocations from the matrix into the precipitates. Classical models treat this contrast through the shear modulus mismatch $\Delta\mu =\mu_{\t{prec}}-\mu_{\t{mat}}$, with strengthening scaling as $|\Delta\mu|^{3/2}$, so that soft and stiff precipitates of equal $|\Delta\mu|$ strengthen equally\autocite{ardell_precipitation_1985}. The dislocation partitioning picture predicts otherwise, since the sign of $\Delta\mu$ decides where dislocations are stored and the three regions do not resist dislocation motion equally (eq.~\ref{eq:region_taylor}). 

Testing this systematically in real alloys is difficult because stiffness cannot be varied alone: changing the precipitate chemistry changes its lattice parameter and interface energy along with it, which both affect strengthening. We circumvent this with computational alchemy. Starting from a BCC matrix, we construct precipitate elements whose stiffness is dialed continuously from much softer to much stiffer than the matrix while their lattice parameter stays matched to it, so the precipitates remain coherent and coherency strains play no role (see Methods section ``\nameref{method:alchemy}''). This is the continuous counterpart of choosing a chemical element for its stiffness, free of the discrete values the periodic table offers and of the solubility limits that decide which element pairs form coherent precipitates at all. Although the alchemical elements represent no real material, the strengthening map they produce spans the full range between soft and stiff precipitates.

Figure \rref{figure_4}{a} shows that strengthening increases with $|\Delta\mu|$ on both sides of the origin $\Delta\mu = 0$, as classical models predict, but its magnitude is not symmetric and depends on the sign of $\Delta \mu$. For example, at $|\Delta\mu|\approx34$ GPa, stiff precipitates ($\Delta\mu>0$) strengthen almost twice as much as soft precipitates ($\Delta\mu<0$), and the shape of the curve differs on the two sides of the origin. Figure~\rref{figure_4}{b} shows the dislocation partitioning behind this asymmetry. Stiff precipitates raise the dislocation density in the matrix and at the interfaces alike, with the precipitates density well below both. Soft precipitates invert this ordering: the highest dislocation density is now inside the precipitates while the matrix is left with the lowest density of all. Equation~\ref{eq:region_taylor} connects this redistribution to strength. The highest densities now sit inside the soft precipitates, where the low shear modulus makes stored dislocations weak obstacles, while the matrix, which occupies most of the volume and controls the flow stress through eq.~\ref{eq:unified_strength}, holds a lower density than it does with stiff precipitates. An obstacle strength that depends on $|\Delta\mu|$ alone, as in the classical models, carries no information about where dislocations are stored and therefore cannot express this redistribution or the asymmetry it produces\autocite{ardell_precipitation_1985, seidman_precipitation_2002, krug_comparison_2014}. 

The strengthening of the Fe-Cr, Cr-Fe, and Fe-Cu samples (all coherent) reported above falls close to the curve computed for the alchemical alloys at comparable $\Delta \mu$ (fig.~\rref{figure_4}{a}), so a real alloy can be located on the map by its stiffness contrast. Figure \rref{figure_4}{a} can thus serve as a practical guide for selecting stiffness contrast.


\section{Discussion and Conclusion}

A key finding of this work is that precipitates actively alter the dislocation density within the matrix itself\autocite{ashby_deformation_1970, queyreau_orowan_2010}. While dislocation storage within precipitates and accumulation at interfaces have direct analogs in classical shearing and bypassing, precipitate-mediated matrix multiplication is not represented in single-dislocation descriptions. This mechanism has been postulated and observed previously\autocite{ashby_deformation_1970, peng_nanoscale_2020, PENG2023103710}, but here we relate it to strengthening and show that it becomes consequential under multi-dislocation conditions. Incorporating multi-dislocation precipitation strengthening into mesoscale descriptions\autocite{bertin_connecting_2019,cui_multiscale_2024} therefore requires accounting for the way precipitates reshape the surrounding dislocation network through coupled shearing, bypassing, and multiplication, in addition to the resistance associated with cutting and bypassing. Existing single-dislocation theories do not represent these concurrent mechanisms, while mesoscale approaches such as discrete dislocation dynamics will require further development to capture them quantitatively.

The comparison between measurements and the classical models in fig.~\rref{figure_1}{e} and Supplementary Information Figure~14c illustrates how neglecting multi-dislocation effects can lead to substantial differences in the predicted strengthening response. Similarly, while precipitate-size polydispersity typically broadens experimental strengthening curves\autocite{seidman_precipitation_2002,fang_statistical_2019,glazer_effect_1988}, it is not a prerequisite for a smooth strengthening peak. Here, a broad crossover emerges intrinsically from the co-evolution of interfacial dislocation accumulation, precipitate storage, and precipitate-mediated dislocation multiplication, even within strictly monodisperse arrays (fig.~\rref{figure_1}{e}). The sign dependence of strengthening on the stiffness contrast (fig.~\ref{figure_4}) reaches into the choice of precipitate chemistry itself. The modulus-mismatch term in the alloy-design models used in practice\autocite{ardell_precipitation_1985, seidman_precipitation_2002, krug_comparison_2014} ranks a soft and a stiff candidate precipitate of equal $|\Delta\mu|$ as equivalent, whereas fig.~\ref{figure_4} shows the stiff one strengthening almost twice as much, because the two place dislocations in different regions of the microstructure. Screening precipitate phases by modulus mismatch alone therefore misranks candidates, and always in the same direction: it overestimates soft precipitates relative to stiff ones.

The observations presented here pertain most closely to deformation regimes characterized by coherent precipitates, high strain rates, and high dislocation densities, where multiple dislocations interact with a single precipitate simultaneously. Such conditions are particularly relevant to additively manufactured alloys and materials subjected to extreme conditions. Alternative conditions --- such as strongly planar slip, semi-coherent or incoherent particles\autocite{yang_shearing_2021}, deformation twinning\autocite{zhong_atomic-scale_2024}, deformation-induced structural transformations\autocite{bourgeois_transforming_2020}, or elevated temperatures --- can activate additional processes (extended pile-ups or climb-assisted relaxation, for example). These processes may introduce, suppress, or redistribute deformation pathways and thereby alter the relative importance of the active plasticity channels (Supplementary Information Notes~9-11). The Fe-Cu system offers one example (Supplementary Information Note~11): deformation-induced transformation of the Cu precipitates adds a fourth channel alongside the three identified here, and classical models, which lack it, overestimate strengthening in the size range where it activates (fig.~\rref{figure_1}{b}). Within its applicability regime, eq.~\ref{eq:unified_strength} still carries a limitation of a different kind: it relates strengthening to a measured dislocation partitioning, and is in that sense descriptive. A predictive theory would have to anticipate the partitioning from the precipitate's own properties, and fig.~\ref{figure_4} is a first step in that direction for the stiffness contrast.

The agreement between the simulated alloys (Fe-Cr, Cr-Fe, and Fe-Cu) and alchemical systems in fig.~\rref{figure_4}{a} suggests that such simulation-derived strengthening maps can guide the choice of precipitate chemistry. Extending them to real alloys demands chemical accuracy and sample size at once, and rapid advances in quantum-accurate machine-learning interatomic potentials\autocite{ko_recent_2023,sheriff2025machinelearningpotentialsmodeling} and increasingly powerful high-performance computing capabilities are now enabling large-scale MD simulations to serve as controlled computational experiments for quantitatively assessing strengthening trends. The present work illustrates what such experiments can deliver: collective mechanisms that are difficult to isolate experimentally, and with them a pathway toward mechanistically informed descriptions of strengthening in precipitate-containing alloys and other heterogeneous microstructures.


\clearpage
\twocolumn[
  \begin{center}
    \Large
    \textbf{Methods}
  \end{center}
]

\subsection{Estimate of dislocations per precipitate}
\label{method:count}
To obtain an order-of-magnitude estimate of the number of dislocations ($N$) that interact with an individual precipitate (fig.~\rref{figure_1}{c}), we start from the bulk dislocation density $\rho$. For a randomly oriented plane, the corresponding effective areal dislocation density is\autocite{spacing}
\[
  \rho_{\t{a}} = \frac{\rho}{2},
\]
which gives the expected number of dislocation lines intersecting that plane for a spatially homogeneous dislocation distribution.

We next estimate a characteristic interaction area associated with a spherical precipitate. For a precipitate of radius $R$, the mean intersection radius with a randomly oriented plane is\autocite{cui_multiscale_2024}
\[
  \bar{R} = \frac{\pi R}{4}.
\]
Dislocations, however, do not need to pass through this strict geometric intersection region to interact with the precipitate. Because the elastic interaction decays approximately as $1/r^3$ with distance $r$ from the precipitate center\autocite{eshelby}, dislocations outside the geometric intersection can still experience appreciable precipitate-induced forces. At a distance equal to twice the geometric intersection radius, the interaction strength is reduced to roughly $(1/2)^3 = 12.5\%$ of its interfacial value. While diminished, this interaction remains non-negligible, implying that dislocations far beyond the strict geometric intersection can still experience appreciable precipitate-induced forces. To provide a conservative and geometrically transparent estimate, we define a characteristic interaction radius
\[
  \bar{R}_{\t{int}} = 2\bar{R},
\]
which captures both the geometric intersection and a physically meaningful elastic interaction range. Here, $\bar{R}_{\t{int}}$ is used only to define the geometric estimate and does not imply that dislocation-precipitate interactions vanish beyond it.

Under these assumptions, the expected number of dislocations interacting
with a precipitate is obtained by multiplying the effective areal
dislocation density by the circular interaction area:
\[
N = \rho_{\mathrm{a}} \pi \bar{R}_{\mathrm{int}}^2
  = \frac{\rho}{2} \pi \left(\frac{\pi R}{2}\right)^2
  = \frac{\pi^3}{8} R^2 \rho.
\]
In fig.~\rref{figure_1}{c}, this expression was evaluated using a representative dislocation density of $\rho = 10^{15}\,\t{m}^{-2}$. Cold-worked metals are commonly reported\autocite{williamson_iii_1956} to exhibit ex situ dislocation densities on the order of $10^{14}\,\t{m}^{-2}$. Such measurements reflect partially relaxed microstructures following deformation, during which processes such as annihilation, junction rearrangement, and recovery reduce the instantaneous dislocation density. During active plastic deformation, transient dislocation densities can exceed the relaxed value\autocite{yang_insitu_2025}. Indeed, substantially higher densities have been measured in heavily deformed engineering alloys, including $8\times10^{15}\,\t{m}^{-2}$ in cold-rolled ODS-Eurofer steel and $1.3\times10^{16}\,\t{m}^{-2}$ in deformation-processed martensitic steel\autocite{renzetti_xray_2012,he_high_2017}. The choice of $\rho = 10^{15}\,\t{m}^{-2}$ therefore represents a reasonable estimate of the instantaneous dislocation density under sustained deformation.

The applicability of this estimate to the samples deformed in this work is discussed in Supplementary Information Note~13.

\subsection{Mechanical deformation}
\label{method:pd}

Molecular dynamics simulations of uniaxial compression were performed on BCC Fe containing spherical Cr precipitates using a timestep of 5\,fs. Atomic interactions between elements were described using an embedded-atom method (EAM) potential\autocite{eich_embedded-atom_2015} that accurately reproduces the Fe-Cr $\alpha/\alpha'$ phase separation and associated thermodynamic and elastic properties. Simulation cells had initial dimensions of $288a_0 \times 288a_0 \times 610a_0$, oriented along the $[100]$, $[010]$, and $[001]$ crystallographic axes, respectively, where $a_0 = 2.86\,\Ang$ is the Fe lattice parameter. Each sample contained approximately $100$ million atoms. Periodic boundary conditions were applied in all three dimensions.

Spherical regions of radius $R$ were placed at random positions within the simulation cell subject to a non-overlap constraint. The number of spheres was chosen to achieve the target precipitate volume fraction $v_{\t{f}}$. All atoms within each sphere were then replaced by Cr atoms, producing coherent spherical BCC Cr precipitates. After precipitate construction, $100$ vacancy-type prismatic dislocation loops of hexagonal shape and 10\,nm diameter were introduced in each sample. The loop orientations were chosen so that all four $\frac{1}{2}\langle 111\rangle$ Burgers vector variants were represented in equal proportion. This procedure yielded an initial dislocation density of $2.55 \times 10^{15}\,\t{m}^{-2}$. Note that the shapes and density of initial sources affect only the transient plastic response whereas steady-state plastic flow does not depend on the exact configuration of initial dislocation sources\autocite{bertin_crystal_2023, bulatov_computer_2006,frank_multiplication_1950, zepeda-ruiz_atomistic_2021, zepeda-ruiz_probing_2017}.

Following loop creation, samples were equilibrated for $100\,\t{ps}$ at $300\,\t{K}$ using a Langevin thermostat (damping parameter $10\,\t{ps}$) and maintained at zero pressure using an isoenthalpic-isobaric (NPH) barostat (damping parameter $10\,\t{ps}$; chain length $3$), allowing only isotropic cell fluctuations. Samples were then deformed in uniaxial compression along $[001]$ at a constant true strain rate\autocite{zepeda-ruiz_atomistic_2021,zepeda-ruiz_probing_2017,zhou_probing_2024}. During deformation, the stresses in the directions orthogonal to the loading axis ($[100]$ and $[010]$) were maintained at zero using an NPH barostat acting on those directions (damping parameter $10\,\t{ps}$; chain length $3$), while the temperature was held at $300\,\t{K}$ using the Langevin thermostat (damping parameter $10\,\t{ps}$).

Most simulations were performed at $\dot{\varepsilon}=10^8\,\t{s}^{-1}$ to a total true strain of $0.6$ (simulation time of $6\,\t{ns}$). For $\dot{\varepsilon}=10^9\,\t{s}^{-1}$, deformation to a true strain of $0.6$ required $0.6\,\t{ns}$. For $\dot{\varepsilon}=10^7\,\t{s}^{-1}$, configurations first deformed at $\dot{\varepsilon}=10^8\,\t{s}^{-1}$ to a true strain of $0.2$ were used as initial states; the strain rate was then reduced to $10^7\,\t{s}^{-1}$ and deformation was continued to a true strain of $0.6$ (simulation time of $40\,\t{ns}$). See Supplementary Information Note~12 for a comparison of strengthening across strain rates.

Reported steady-state flow stresses are obtained by averaging flow stress over the true-strain interval $\varepsilon = 0.4-0.6$; error bars denote autocorrelation-corrected standard errors of the mean. Within this interval, the dislocation networks remain statistically stationary (Supplementary Information Note~4), making the steady-state flow stress a robust atomistic measure of material strength\autocite{zepeda-ruiz_atomistic_2021, zepeda-ruiz_probing_2017}. This steady-state flow stress differs from the yield strength measured in experiments and predicted by the classical models (figs.~\rref{figure_1}{b} and \rref{figure_1}{e}): the yield strength marks the onset of plastic flow, whereas the steady-state flow stress is the stress required to sustain plastic flow against a fully developed dislocation network. Both quantities, however, increase with the resistance that precipitates offer to dislocation motion, and the precipitation strengthening $\Delta\sigma$ isolates this precipitate contribution by subtracting the response of the precipitate-free matrix deformed under identical conditions. The comparison in figs.~\rref{figure_1}{b} and \rref{figure_1}{e} therefore concerns the dependence of $\Delta\sigma$ on precipitate size, i.e., the position of the strengthening maximum and the breadth of the crossover between the shearing and bypass regimes, which is set by the dislocation-precipitate interactions and should therefore hold for both measures of strength.

All simulations were carried out using the Large-scale Atomic/Molecular Massively Parallel Simulator (LAMMPS)\autocite{thompson_lammps_2022} and visualizations were done using Open Visualization Tool \autocite{stukowski_visualization_2009} (OVITO).

\subsection{Composition-based voxel segmentation of dislocation}
\label{method:dxa}

Dislocations were identified using the Dislocation Extraction Algorithm (DXA)\autocite{stukowski_extracting_2010}. To quantify spatial variations in dislocation density, each snapshot was discretized into cubic voxels of size $6\,\Ang \times 6\,\Ang \times 6\,\Ang$. A voxel size of $6\,\Ang$ exceeds twice the BCC Fe lattice parameter and encompasses interactions up to the sixth nearest-neighbor shell, providing adequate atomic sampling ($\approx 20$ atoms) while maintaining the spatial resolution required to resolve precipitates as small as $R \approx 7\,\Ang$. The local Cr atomic fraction $c_{\t{Cr}}$ was computed for each voxel and used to classify voxels as matrix ($c_{\t{Cr}} \le 0.1$), interface ($0.1 < c_{\t{Cr}} < 0.9$), or precipitate ($0.9 \le c_{\t{Cr}}$). The microstructure fractions of interface ($f_{\t{int}}$) and precipitate ($f_{\t{prec}}$) were computed as the fraction of voxels belonging to each region. Importantly, $f_{\t{prec}}$ differs from the nominal precipitate volume fraction $v_\t{f}$: $f_{\t{prec}}$ quantifies the volume fraction of precipitate regions at steady state as evaluated by our composition-based voxel segmentation method, whereas $v_\t{f}$ denotes the initial prescribed precipitate volume fraction based on the ratio of Cr to Fe atoms. 

For each region, the associated dislocation density was calculated by summing the total dislocation line length contained within voxels of that region\autocite{bertin_connecting_2019} and normalizing by the corresponding region volume. Reported steady-state dislocation properties and region fractions are averaged over the true-strain interval $\varepsilon=0.4-0.6$ and error bars denote autocorrelation-corrected standard errors of the mean.

Note that the precipitate and interface regions can be smaller than the characteristic forest spacing implied by eq.~\ref{eq:region_taylor}. Accordingly, the computed dislocation densities $\rho_\t{prec}$ and $\rho_\t{int}$ using the aforementioned methodology should not be interpreted as local forest densities within an individual precipitate or interface region. Rather, they represent ensemble-averaged dislocation densities over the full population of precipitate and interface regions.

\subsection{Precipitate strain evaluation}
\label{method:prec_strain}

The precipitate strain at the end of deformation was computed from precipitate shapes obtained by cluster analysis. For each cluster (i.e., precipitate), we calculated its gyration tensor $\mathbf{G}$. The precipitate extent along the loading direction $[001]$ was defined from the projection of $\mathbf{G}$ onto the unit vector $\hat{\mathbf{e}}_{[001]}$ as\autocite{mattice_conformational_1994}
\[
\ell_{\t{p}}^{2} = \hat{\mathbf{e}}_{[001]}^{\mathsf{T}}\, \mathbf{G}\, \hat{\mathbf{e}}_{[001]} .
\]
The true strain of a precipitate along $[001]$ was then defined as,
\[
\varepsilon_{\t{prec}} = -\ln\!\left(\frac{\ell_{\t{p}}}{\ell_{\t{p},0}}\right),
\]
where $\ell_{\t{p},0}$ is the corresponding extent measured from the undeformed configuration of the same precipitate. The reported precipitate strain $\bar{\varepsilon}_{\t{prec}}$ was obtained by averaging $\varepsilon_{\t{prec}}$ over all precipitates in the sample.

\subsection{Partitioning-based strengthening relation}
\label{method:model}

The strengthening predicted by the partitioning-based relation, shown on the x-axis of fig.~\ref{figure_3}, was evaluated using eq.~\ref{eq:unified_strength}. The required dislocation densities and volume fractions, $\rho_{\t{mat}}$, $\rho_{\t{int}}$, $\rho_{\t{prec}}$, $f_{\t{mat}}$, $f_{\t{int}}$, and $f_{\t{prec}}$, were obtained directly from the MD simulations using the composition-resolved dislocation segmentation procedure described in \nameref{method:dxa}.

The region-resolved stress contributions, $\sigma_{\t{mat}}$, $\sigma_{\t{int}}$, and $\sigma_{\t{prec}}$, were computed using eq.~\ref{eq:region_taylor}. For each material system, the shear modulus and Burgers vector of the matrix and precipitate regions were computed from the corresponding perfect crystals at $0\,\t{K}$. The interfacial shear modulus and Burgers vector were then taken as the arithmetic averages of the corresponding matrix and precipitate values.

For each material system, the region-resolved Taylor coefficients $\alpha_{\t{mat}}$, $\alpha_{\t{int}}$, and $\alpha_{\t{prec}}$ were obtained by calibrating eq.~\ref{eq:unified_strength} against the flow stresses measured from the corresponding MD simulations. Once calibrated, the same set of coefficients was used for all microstructures within a given material system.

\subsection{Alchemical model element construction}
\label{method:alchemy}

To probe the effect of modulus mismatch on precipitation strengthening, we adopted the computational alchemy methodology of ref.~\cite{liang_computational_2025}. The Zhou et al.\autocite{zhou_potential} EAM parameterization for Ta was used as the reference matrix potential. To generate model precipitate elements with controlled stiffness, all $11$ energy-dimensional parameters of the Ta potential were scaled by a factor $k$, producing an element with stiffness $k$ times that of the matrix Ta. This procedure allowed us to construct alchemical matrix-precipitate alloy models in which the modulus mismatch could be systematically varied while retaining the same underlying crystal structure and lattice compatibility.

Elastic modulus is not the only material property affected by the rescaling of energy-dimensional parameters. Cohesive energy, defect energetics, Peierls barriers to dislocation motion, and melting point are all altered as well\autocite{liang_computational_2025}. This is because, like the elastic response, they arise from the same underlying interatomic energy landscape and therefore naturally covary with stiffness as that landscape changes. Indeed, within the same computational-alchemy framework, the unstable stacking-fault energy was found to vary nearly linearly with stiffness\autocite{liang_computational_2025}. We find a similar nearly linear dependence of melting point on stiffness (Supplementary Information Figure~17a). In real alloys, the same physical coupling persists: changes in stiffness are accompanied by changes in other properties governed by the underlying energy landscape. Thus, we deliberately retain this coupled variation as part of the strengthening response associated with changing precipitate stiffness in our alchemical models, rather than artificially constraining the elastic modulus to vary independently of other energy-derived properties.

Seven values of $k$ from $0.1$ to $3.0$ were considered. All values produced coherent BCC precipitates, except for $k=0.1$, for which some precipitates became semi-coherent or incoherent during deformation. This loss of coherency is consistent with the low melting point of the ($k=0.1$) precipitates ($\sim330\,\t{K}$), which is close to the simulation temperature ($300\,\t{K}$). However, the strengthening response in fig.~\rref{figure_4}{a} remains asymmetric even when the $k=0.1$ case is excluded and when the simulations are repeated at $50\,\t{K}$, indicating that the asymmetry is not simply a consequence of differences in homologous temperature (see Supplementary Information Note 14). The modulus mismatch reported in fig.~\ref{figure_4} was defined as $\Delta\mu=\mu_{\t{prec}}-\mu_{\t{mat}}$, where $\mu_{\t{prec}}$ and $\mu_{\t{mat}}$ are the Voigt-averaged shear moduli of the corresponding model precipitate and Ta matrix elements~\autocite{BERRYMAN20053730}.

Using Ta as the matrix element and model alchemical elements as the precipitate elements, we performed MD simulations of mechanical deformation at $300\,\t{K}$ and a strain rate of $10^8~\t{s}^{-1}$. Each system contained approximately $100$ million atoms and was constructed with a lattice constant of $3.303~\Ang$. The simulations followed the same procedure described in \nameref{method:pd}. Dislocation partitioning was analyzed using the procedure described in \nameref{method:dxa}, except that a voxel size of $7\,\Ang \times 7\,\Ang \times 7\,\Ang$ was used.

\subsection{Data availability}

The raw simulation data have not been deposited in a public repository because of their large size, which spans multiple terabytes. Processed data sufficient to reproduce all figures and results in the paper and Supplementary Information will be made available upon publication in a public repository.

\subsection{Code availability}
   
LAMMPS software\autocite{thompson_lammps_2022} used in all molecular dynamics simulations reported in this work is available at \texttt{https: //www.lammps.org}. Our figure style is implemented in LovelyPlots\autocite{sheriff_lovelyplots_2022} under the paper style. Code to reproduce all figures in the paper from processed data will be made available upon publication in a public repository. Any custom code that is not currently available can be subsequently added upon reasonable request to the corresponding author.

\subsection{Acknowledgments}

This material is based upon work supported by the Air Force Office of Scientific Research (AFOSR) under award number FA9550-25-1-0199, through the Young Investigator Program. This work is also supported by the MathWorks Ignition Fund. M.I. and V.V.B. acknowledge funding support from the Laboratory Directed Research and Development Program.  All authors acknowledge computational time allocation on Tuolumne supercomputer from the Computational Grand Challenge program at Lawrence Livermore National Laboratory. This work was performed under the auspices of the U.S. Department of Energy by Lawrence Livermore National Laboratory, USA under Contract DE-AC52-07NA27344.

\subsection{Author contributions statement}

M.I., V.V.B., and R.F. conceived the project.
M.I. performed molecular dynamics simulations and analyzed the results.
J.O. performed melting-point calculations.
All authors contributed to the interpretation of the results, prepared, reviewed, and edited the manuscript.
R.F. and V.V.B. supervised the project.  

\subsection{Competing interests statement}

The authors declare no competing interests.

\clearpage
\printbibliography[heading=bibnumbered,title={References}]

@book{kelly_strengthening_1971,
	title = {Strengthening methods in crystals},
	isbn = {978-0-444-20105-8},
	publisher = {Amsterdam, New York, Elsevier Pub. Co.},
	author = {Kelly, A. (Anthony)},
	date = {1971},
}

@book{argon_strengthening_2007,
	title = {Strengthening Mechanisms in Crystal Plasticity},
	isbn = {978-0-19-851600-2},
	publisher = {Oxford University Press},
	author = {Argon, Ali},
	date = {2007-08-30},
	doi = {10.1093/acprof:oso/9780198516002.001.0001},
}

@article{reppich_new_1982,
	title = {Some new aspects concerning particle hardening mechanisms in {\ensuremath{\gamma'}} precipitating Ni-base alloys—I. Theoretical concept},
	issn = {0001-6160},
	doi = {10.1016/0001-6160(82)90048-7},
	journaltitle = {Acta Metallurgica},
	author = {Reppich, Bernd},
	date = {1982-01-01},
}

@article{ardell_precipitation_1985,
	title = {Precipitation hardening},
	volume = {16},
	issn = {2379-0180},
	doi = {10.1007/BF02670416},
	journaltitle = {Metallurgical Transactions A},
	author = {Ardell, A. J.},
	date = {1985-12-01},
}

@article{cui_multiscale_2024,
	title = {Multiscale modelling of precipitation hardening: a review},
	issn = {3004-8966},
	doi = {10.1186/s41313-024-00066-6},
	journaltitle = {Journal of Materials Science: Materials Theory},
	author = {Cui, Aiya and Wang, Xiaoming and Cui, Yinan},
	date = {2024-07-16},
}

@article{orowan1948symposium,
  title={Symposium on internal stresses in metals and alloys},
  author={Orowan, E},
  journal={Institute of Metals, London},
  volume={451},
  year={1948}
}

@article{jiang_ultrastrong_2017,
	title = {Ultrastrong steel via minimal lattice misfit and high-density nanoprecipitation},
	doi = {10.1038/nature22032},
	journaltitle = {Nature},
	author = {Jiang, Suihe and Wang, Hui and Wu, Yuan and Liu, Xiongjun and Chen, Honghong and Yao, Mengji and Gault, Baptiste and Ponge, Dirk and Raabe, Dierk and Hirata, Akihiko and Chen, Mingwei and Wang, Yandong and Lu, Zhaoping},
	date = {2017-04},
}

@article{jang_shear_2021,
	title = {Shear band-driven precipitate dispersion for ultrastrong ductile medium-entropy alloys},
	doi = {10.1038/s41467-021-25031-6},
	journaltitle = {Nature Communications},
	author = {Jang, Tae Jin and Choi, Won Seok and Kim, Dae Woong and Choi, Gwanghyo and Jun, Hosun and Ferrari, Alberto and Körmann, Fritz and Choi, Pyuck-Pa and Sohn, Seok Su},
	date = {2021-08-04},
}

@article{tang_precipitation_2019,
	title = {Precipitation strengthening in an ultralight magnesium alloy},
	doi = {10.1038/s41467-019-08954-z},
	journaltitle = {Nature Communications},
	author = {Tang, Song and Xin, Tongzheng and Xu, Wanqiang and Miskovic, David and Sha, Gang and Quadir, Zakaria and Ringer, Simon and Nomoto, Keita and Birbilis, Nick and Ferry, Michael},
	date = {2019-03-01},
}

@article{sun_precipitation_2019,
	title = {Precipitation strengthening of aluminum alloys by room-temperature cyclic plasticity},
	doi = {10.1126/science.aav7086},
	journaltitle = {Science},
	author = {Sun, Wenwen and Zhu, Yuman and Marceau, Ross and Wang, Lingyu and Zhang, Qi and Gao, Xiang and Hutchinson, Christopher},
	date = {2019},
}

@article{wang_shearing_2024,
	title = {Shearing brittle intermetallics enhances cryogenic strength and ductility of steels},
	doi = {10.1126/science.ado2919},
	journaltitle = {Science},
	author = {Wang, Feng and Song, Miao and Elkot, Mohamed N. and Yao, Ning and Sun, Binhan and Song, Min and Wang, Zhangwei and Raabe, Dierk},
	date = {2024-05-31},
}

@article{seidman_precipitation_2002,
	title = {Precipitation strengthening at ambient and elevated temperatures of heat-treatable Al(Sc) alloys},
	doi = {10.1016/S1359-6454(02)00201-X},
	journaltitle = {Acta Materialia},
	author = {Seidman, David N. and Marquis, Emmanuelle A. and Dunand, David C.},
	date = {2002-09-20},
}

@article{osamura_precipitation_1994,
	title = {Precipitation Hardening in Fe-Cu Binary and Quaternary Alloys},
	doi = {10.2355/isijinternational.34.359},
	journaltitle = {{ISIJ} International},
	author = {Osamura, Kozo and Okuda, Hiroshi and Ochiai, Shojiro and Takashima, Minoru and Asano, Kazuo and Furusaka, Michihiro and Kishida, Koji and Kurosawa, Fumio},
	date = {1994},
}

@article{zepeda-ruiz_probing_2017,
	title        = {Probing the limits of metal plasticity with molecular dynamics simulations},
	author       = {Zepeda-Ruiz, Luis A. and Stukowski, Alexander and Oppelstrup, Tomas and Bulatov, Vasily V.},
	doi          = {10.1038/nature23472},
	journaltitle = {Nature},
	date         = {2017-10},
}

@article{zepeda-ruiz_atomistic_2021,
	title        = {Atomistic insights into metal hardening},
	author       = {Zepeda-Ruiz, Luis A. and Stukowski, Alexander and Oppelstrup, Tomas and Bertin, Nicolas and Barton, Nathan R. and Freitas, Rodrigo and Bulatov, Vasily V.},
	doi          = {10.1038/s41563-020-00815-1},
	journaltitle = {Nature Materials},
	date         = {2021-03},
}

@article{zhou_probing_2024,
	title = {Probing multi-dimensional composition spaces in search of strong metallic alloys},
	doi = {10.1038/s41524-026-01975-5},
	journaltitle = {npj Computational Materials},
	author = {Zhou, Xinran and Marian, Jaime and Zhou, Fei and Bulatov, Vasily V.},
	date = {2026-01-31},
}

@article{van_dalen_microstructural_2011,
	title = {Microstructural evolution and creep properties of precipitation-strengthened Al–0.06Sc–0.02Gd and Al–0.06Sc–0.02Yb (at.\%) alloys},
	doi = {10.1016/j.actamat.2011.04.059},
	journaltitle = {Acta Materialia},
	author = {Van Dalen, Marsha E. and Dunand, David C. and Seidman, David N.},
	date = {2011-08},
}

@article{islam_nonequilibrium_2025,
	title = {Nonequilibrium chemical short-range order in metallic alloys},
	doi = {10.1038/s41467-025-64733-z},
	journaltitle = {Nature Communications},
	author = {Islam, Mahmudul and Sheriff, Killian and Cao, Yifan and Freitas, Rodrigo},
	date = {2025-10-08},
}

@article{islam_dislocation-mediated_2025,
	title = {Dislocation-mediated short-range order evolution during thermomechanical processing},
	doi = {https://doi.org/10.1016/j.actamat.2025.121838},
	journaltitle = {Acta Materialia},
	author = {Islam, Mahmudul and Sheriff, Killian and Freitas, Rodrigo},
	date = {2025},
}

@article{takahashi_computational_2008,
	title = {A computational method for dislocation–precipitate interaction},
	doi = {10.1016/j.jmps.2007.08.002},
	journaltitle = {Journal of the Mechanics and Physics of Solids},
	author = {Takahashi, Akiyuki and Ghoniem, Nasr M.},
	date = {2008-04-01},
}

@article{peng_nanoscale_2020,
	title = {Nanoscale precipitates as sustainable dislocation sources for enhanced ductility and high strength},
	doi = {10.1073/pnas.1914615117},
	journaltitle = {Proceedings of the National Academy of Sciences},
	author = {Peng, Shenyou and Wei, Yujie and Gao, Huajian},
	date = {2020-03-10},
}

@article{krug_comparison_2014,
	title = {Comparison between dislocation dynamics model predictions and experiments in precipitation-strengthened Al–Li–Sc alloys},
	doi = {10.1016/j.actamat.2014.06.038},
	journaltitle = {Acta Materialia},
	author = {Krug, Matthew E. and Mao, Zugang and Seidman, David N. and Dunand, David C.},
	date = {2014-10},
}

@book{nembach_particle_1997,
	title = {Particle strengthening of metals and alloys},
	isbn = {0-471-12072-3},
	publisher = {Wiley},
	author = {Nembach, E.},
	date = {1997},
}

@article{fang_statistical_2019,
	title = {A statistical theory of probability-dependent precipitation strengthening in metals and alloys},
	doi = {10.1016/j.jmps.2018.09.010},
	journaltitle = {Journal of the Mechanics and Physics of Solids},
	author = {Fang, Qihong and Li, Li and Li, Jia and Wu, Hongyu and Huang, Zaiwang and Liu, Bin and Liu, Yong and Liaw, Peter K},
	date = {2019-01-01},
}

@article{glazer_effect_1988,
	title = {The effect of the precipitate size distribution on the aging curve of order hardening alloys},
	doi = {10.1016/0001-6160(88)90145-9},
	journaltitle = {Acta Metallurgica},
	author = {Glazer, J. and Morris, J. W.},
	date = {1988-04-01},
}

@article{liang_computational_2025,
	title = {Computational alchemy clarifies origins of alloy strengthening},
	doi = {10.1038/s41524-025-01910-0},
	journaltitle = {npj Computational Materials},
	author = {Liang, Aoyan and Bertin, Nicolas and Zhou, Xinran and Aubry, Sylvie and Bulatov, Vasily V.},
	date = {2025-12-18},
}

@article{ashby_deformation_1970,
	title = {The deformation of plastically non-homogeneous materials},
	doi = {10.1080/14786437008238426},
	journaltitle = {The Philosophical Magazine: A Journal of Theoretical Experimental and Applied Physics},
	author = {Ashby, M. F.},
	date = {1970-02-01},
}

@article{bulatov_network_2025,
	title = {Network aspects of single crystal plasticity},
	doi = {10.1016/j.jmrt.2025.04.014},
	journaltitle = {Journal of Materials Research and Technology},
	author = {Bulatov, Vasily V. and Bertin, Nicolas and Aubry, Sylvie and Zepeda-Ruiz, Luis A. and Zhou, Xinran and Liang, Aoyan and Oppelstrup, Tomas and Sadigh, Babak and Arsenlis, Athanasios and Austin, Ryan and Carson, Robert and Barton, Nathan R.},
	date = {2025-05-01},
}

@article{taylor_mechanism_1934,
	title = {The mechanism of plastic deformation of crystals. Part I.—Theoretical},
	doi = {10.1098/rspa.1934.0106},
	journaltitle = {Proceedings of the Royal Society of London. Series A, Containing Papers of a Mathematical and Physical Character},
	author = {Taylor, Geoffrey Ingram},
	date = {1934-07-02},
}

@article{zhu_heterostructured_2023,
	title = {Heterostructured materials},
	doi = {10.1016/j.pmatsci.2022.101019},
	journaltitle = {Progress in Materials Science},
	author = {Zhu, Yuntian and Wu, Xiaolei},
	date = {2023-01-01},
}

@article{picu_superposition_2010,
	title = {On the superposition of flow stress contributions at finite temperatures and in the athermal limit},
	doi = {10.1016/j.actamat.2010.06.020},
	journaltitle = {Acta Materialia},
	author = {Picu, R.C. and Li, R.},
	date = {2010-09},
}

@article{williamson_iii_1956,
	title = {{III}. Dislocation densities in some annealed and cold-worked metals from measurements on the X-ray debye-scherrer spectrum},
	doi = {10.1080/14786435608238074},
	journaltitle = {The Philosophical Magazine: A Journal of Theoretical Experimental and Applied Physics},
	author = {Williamson, G. K. and Smallman, R. E.},
	date = {1956-01-01},
}

@article{kovarik_microtwinning_2009,
	title = {Microtwinning and other shearing mechanisms at intermediate temperatures in Ni-based superalloys},
	doi = {10.1016/j.pmatsci.2009.03.010},
	journaltitle = {Progress in Materials Science},
	author = {Kovarik, L. and Unocic, R. R. and Li, Ju and Sarosi, P. and Shen, C. and Wang, Y. and Mills, M. J.},
	date = {2009-08-01},
}

@software{sheriff_lovelyplots_2022,
  author       = {Killian Sheriff},
  title        = {{LovelyPlots, a collection of matplotlib 
                   stylesheets for scientific figures}},
  month        = jul,
  year         = 2022,
  publisher    = {Zenodo},
  doi          = {10.5281/zenodo.6903936},
}

@article{zhong_atomic-scale_2024,
	title = {Atomic-scale observation of nucleation- and growth-controlled deformation twinning in body-centered cubic nanocrystals},
	doi = {10.1038/s41467-024-44837-8},
	journaltitle = {Nature Communications},
	author = {Zhong, Li and Zhang, Yin and Wang, Xiang and Zhu, Ting and Mao, Scott X.},
	date = {2024-01-16},
}

@article{bourgeois_transforming_2020,
	title = {Transforming solid-state precipitates via excess vacancies},
	doi = {10.1038/s41467-020-15087-1},
	journaltitle = {Nature Communications},
	author = {Bourgeois, Laure and Zhang, Yong and Zhang, Zezhong and Chen, Yiqiang and Medhekar, Nikhil V.},
	date = {2020-03-06},
}

@article{stukowski_extracting_2010,
	title        = {Extracting dislocations and non-dislocation crystal defects from atomistic simulation data},
	author       = {Stukowski, Alexander and Albe, Karsten},
	doi          = {10.1088/0965-0393/18/8/085001},
	issn         = {0965-0393},
	journaltitle = {Modelling and Simulation in Materials Science and Engineering},
	date         = {2010-09},
}

@article{stukowski_visualization_2009,
	title        = {Visualization and analysis of atomistic simulation data with {OVITO}–the Open Visualization Tool},
	author       = {Stukowski, Alexander},
	doi          = {10.1088/0965-0393/18/1/015012},
	issn         = {0965-0393},
	journaltitle = {Modelling and Simulation in Materials Science and Engineering},
	date         = {2009-12},
}

@article{thompson_lammps_2022,
	title = {{LAMMPS} - a flexible simulation tool for particle-based materials modeling at the atomic, meso, and continuum scales},
	doi = {10.1016/j.cpc.2021.108171},
	journaltitle = {Computer Physics Communications},
	author = {Thompson, Aidan P. and Aktulga, H. Metin and Berger, Richard and Bolintineanu, Dan S. and Brown, W. Michael and Crozier, Paul S. and in 't Veld, Pieter J. and Kohlmeyer, Axel and Moore, Stan G. and Nguyen, Trung Dac and Shan, Ray and Stevens, Mark J. and Tranchida, Julien and Trott, Christian and Plimpton, Steven J.},
	date = {2022-02-01},
}

@article{eich_embedded-atom_2015,
	title = {Embedded-atom potential for an accurate thermodynamic description of the iron–chromium system},
	doi = {10.1016/j.commatsci.2015.03.047},
	journaltitle = {Computational Materials Science},
	author = {Eich, S. M. and Beinke, D. and Schmitz, G.},
	date = {2015-06-15},
}

@article{bertin_connecting_2019,
	title = {Connecting discrete and continuum dislocation mechanics: A non-singular spectral framework},
	doi = {10.1016/j.ijplas.2018.12.006},
	journaltitle = {International Journal of Plasticity},
	author = {Bertin, Nicolas},
	date = {2019-11-01},
}

@article{bertin_crystal_2023,
	title = {Crystal plasticity model of {BCC} metals from large-scale {MD} simulations},
	doi = {10.1016/j.actamat.2023.119336},
	journaltitle = {Acta Materialia},
	author = {Bertin, Nicolas and Carson, Robert and Bulatov, Vasily V. and Lind, Jonathan and Nelms, Matthew},
	date = {2023-11-01},
}

@book{bulatov_computer_2006,
	title = {Computer Simulations of Dislocations},
	isbn = {978-0-19-852614-8},
	publisher = {{OUP} Oxford},
	author = {Bulatov, Vasily and Cai, Wei},
	date = {2006-11-02},
}

@article{frank_multiplication_1950,
	title = {Multiplication Processes for Slow Moving Dislocations},
	doi = {10.1103/PhysRev.79.722},
	journaltitle = {Physical Review},
	author = {Frank, F. C. and Read, W. T.},
	date = {1950-08-15},
}

@book{mattice_conformational_1994,
	title = {Conformational theory of large molecules : the rotational isomeric state model in macromolecular systems},
	publisher = {Wiley},
	author = {Mattice, Wayne L. and Suter, U.},
	date = {1994},
}

@article{zhou_designing_2026,
	title = {Designing heterostructured materials},
	doi = {10.1038/s41563-025-02444-y},
	journaltitle = {Nature Materials},
	author = {Zhou, Hao and Wu, Xiaolei and Srolovitz, David and Zhu, Yuntian},
	date = {2026-01-13},
}

@article{wilm1911physical,
  title        = {Physical metallurgical experiments on aluminum alloys containing magnesium},
  author       = {Wilm, A.},
  journal      = {Metallurgie},
  year         = {1911},
}

@article{spacing,
    author = {Schoeck, G.},
    title = {Correlation between Dislocation Length and Density},
    journal = {Journal of Applied Physics},
    year = {1962},
    doi = {10.1063/1.1728821},
}

@article{tiley_coarsening_2009,
	title = {Coarsening kinetics of $\gamma '$ precipitates in the commercial nickel base Superalloy René 88 {DT}},
	doi = {10.1016/j.actamat.2009.02.010},
	journaltitle = {Acta Materialia},
	author = {Tiley, J. and Viswanathan, G. B. and Srinivasan, R. and Banerjee, R. and Dimiduk, D. M. and Fraser, H. L.},
	date = {2009-05-01},
}

@article{baker_processes_2009,
	title = {Processes, microstructure and properties of vanadium microalloyed steels},
	doi = {10.1179/174328409X453253},
	journaltitle = {Materials Science and Technology},
	author = {Baker, T. N.},
	date = {2009},
}

@article{lee_evaluation_2023,
	title = {Evaluation of precipitation phase and mechanical properties according to aging heat treatment temperature of Inconel 718},
	doi = {10.1016/j.jmrt.2023.10.196},
	journaltitle = {Journal of Materials Research and Technology},
	author = {Lee, Gang Ho and Park, Minha and Kim, Byoungkoo and Jeon, Jong Bae and Noh, Sanghoon and Kim, Byung Jun},
	date = {2023-11-01},
}

@article{lee_precipitate_2025,
	title = {Precipitate phase behavior and mechanical properties of Inconel 718 according to aging heat treatment time},
	doi = {10.1016/j.msea.2024.147776},
	journaltitle = {Materials Science and Engineering: A},
	author = {Lee, Gang Ho and Kim, Byoungkoo and Jeon, Jong Bae and Park, Minha and Noh, Sanghoon and Kim, Byung Jun},
	date = {2025-02-01},
}

@report{mohale_microstructural_2022,
	title = {Microstructural Characterization of A709 Commercial Heats with Precipitation Treatment},
	number = {{INL}/{RPT}-22-68666-Rev000},
	institution = {Idaho National Laboratory ({INL}), Idaho Falls, {ID} (United States)},
	author = {Mohale, Ninad and Bass, Ryann Elizabeth},
	date = {2022-08-16},
}

@article{seltzman_precipitate_2023,
	title = {Precipitate Size in GRCop-42 and GRCop-84 Cu-Cr-Nb Alloy Gas Atomized Powder and L-PBF Additive Manufactured Material},
	doi = {10.1080/15361055.2022.2147765},
	journaltitle = {Fusion Science and Technology},
	author = {Seltzman, A. H. and Wukitch, S. J.},
	date = {2023},
}

@article{eshelby,
    author = {Eshelby, John Douglas},
    title = {The determination of the elastic field of an ellipsoidal inclusion, and related problems},
    journal = {Proceedings of the Royal Society of London. A. Mathematical and Physical Sciences},
    date = {1957},
    doi = {10.1098/rspa.1957.0133},
}

@article{PENG2023103710,
title = {Beyond Orowan hardening: Mapping the four distinct mechanisms associated with dislocation-precipitate interaction},
journal = {International Journal of Plasticity},
year = {2023},
doi = {https://doi.org/10.1016/j.ijplas.2023.103710},
author = {Shenyou Peng and Zhili Wang and Jia Li and Qihong Fang and Yujie Wei},
}

@article{LAGERPUSCH20003647,
title = {Double strengthening of copper by dissolved gold-atoms and by incoherent SiO2-particles: how do the two strengthening contributions superimpose?},
journal = {Acta Materialia},
year = {2000},
doi = {https://doi.org/10.1016/S1359-6454(00)00172-5},
author = {U Lagerpusch and V Mohles and D Baither and B Anczykowski and E Nembach},
}

@article{fan_strain_2021,
	title = {Strain rate dependency of dislocation plasticity},
	doi = {10.1038/s41467-021-21939-1},
	journaltitle = {Nature Communications},
	publisher = {Nature Publishing Group},
	author = {Fan, Haidong and Wang, Qingyuan and El-Awady, Jaafar A. and Raabe, Dierk and Zaiser, Michael},
	date = {2021-03-23},
}

@article{ko_recent_2023,
	title = {Recent advances and outstanding challenges for machine learning interatomic potentials},
	doi = {10.1038/s43588-023-00561-9},
	journaltitle = {Nature Computational Science},
	author = {Ko, Tsz Wai and Ong, Shyue Ping},
	date = {2023-12},
}

@article{zhou_potential,
  title = {Misfit-energy-increasing dislocations in vapor-deposited CoFe/NiFe multilayers},
  author = {Zhou, X. W. and Johnson, R. A. and Wadley, H. N. G.},
  journal = {Phys. Rev. B},
  year = {2004},
  doi = {10.1103/PhysRevB.69.144113},
}

@article{BERRYMAN20053730,
title = {Bounds and estimates for elastic constants of random polycrystals of laminates},
journal = {International Journal of Solids and Structures},
year = {2005},
doi = {https://doi.org/10.1016/j.ijsolstr.2004.12.015},
author = {James G. Berryman},
}

@misc{sheriff2025machinelearningpotentialsmodeling,
      title={Machine learning potentials for modeling alloys across compositions}, 
      author={Killian Sheriff and Daniel Xiao and Yifan Cao and Lewis R. Owen and Rodrigo Freitas},
      year={2025},
      doi={10.48550/arXiv.2506.12592},
}

@article{chatterjee_discrete_2021,
	title = {A discrete dislocation dynamics study of precipitate bypass mechanisms in nickel-based superalloys},
	doi = {10.1016/j.ijplas.2021.103062},
	journaltitle = {International Journal of Plasticity},
	author = {Chatterjee, Sabyasachi and Li, Yang and Po, Giacomo},
	date = {2021-10-01},
}

@article{stewart_predicting_2024,
	title = {Predicting yield stress in a nano-precipitate strengthened austenitic steel by integrating multi length-scale simulations and experiments},
	doi = {10.1016/j.actamat.2024.119918},
	journaltitle = {Acta Materialia},
	author = {Stewart, Colin A. and Antillon, Edwin A. and Sudmanns, Markus and El-Awady, Jaafar A. and Knipling, Keith E. and Callahan, Patrick G. and Rowenhorst, David J. and Fonda, Richard W.},
	date = {2024-06-15},
}

@article{queyreau_orowan_2010,
	title = {Orowan strengthening and forest hardening superposition examined by dislocation dynamics simulations},
	doi = {10.1016/j.actamat.2010.06.028},
	journaltitle = {Acta Materialia},
	author = {Queyreau, Sylvain and Monnet, Ghiath and Devincre, Benoit},
	date = {2010-10-01},
}

@article{yang_shearing_2021,
	title = {Shearing and rotation of $\beta$" and $\beta$' precipitates in an Al-Mg-Si alloy under tensile deformation: \textit{In-situ} and \textit{ex-situ} studies},
	doi = {10.1016/j.actamat.2021.117310},
	journaltitle = {Acta Materialia},
	author = {Yang, Mingjun and Orekhov, Andrey and Hu, Zhi-Yi and Feng, Man and Jin, Shenbao and Sha, Gang and Li, Kai and Samaee, Vahid and Song, Min and Du, Yong and Van Tendeloo, Gustaaf and Schryvers, Dominique},
	date = {2021-11-01},
}

@article{he_high_2017,
  author  = {He, B. B. and Hu, B. and Yen, H. W. and Cheng, G. J. and Wang, Z. K. and Luo, H. W. and Huang, M. X.},
  title   = {High dislocation density-induced large ductility in deformed and partitioned steels},
  journal = {Science},
  year    = {2017},
  doi     = {10.1126/science.aan0177}
}

@article{yang_insitu_2025,
  author  = {Yang, Zhening and Thomas, Rhys and Donoghue, Jack and Smith, Albert and Preuss, Michael and Gholinia, Ali},
  title   = {In situ and correlative study of dislocation density and deformation mechanisms in {Inconel 690}},
  journal = {Materials Characterization},
  year    = {2025},
  doi     = {10.1016/j.matchar.2025.115699}
}

@article{renzetti_xray_2012,
  author  = {Renzetti, R. A. and Sandim, H. R. Z. and Bolmaro, R. E. and Suzuki, P. A. and M{\"o}slang, A.},
  title   = {X-ray evaluation of dislocation density in {ODS-Eurofer} steel},
  journal = {Materials Science and Engineering: A},
  year    = {2012},
  doi     = {10.1016/j.msea.2011.11.051}
}

\end{document}